\documentclass[margin=0.5in]{article}

\usepackage[margin=1in]{geometry}

\usepackage{colortbl}
\usepackage{adjustbox}
\newcommand*\rot{\rotatebox{90}}
\usepackage{multirow}
\usepackage{arydshln}
\usepackage{cite}
\usepackage{bbding}
\usepackage{dingbat}
\usepackage[normalem]{ulem}

\usepackage{hyperref}

\date{}

\usepackage{cleveref}
\crefname{figure}{fig.}{figs.} % Needed for ieee style abbrev (Fig.) % https://tex.stackexchange.com/a/609905

\newcommand{\peerreviewhide}[1]{{ }}

\newcommand{\cg}[1]{{\color{black}{#1}}}
\newcommand{\jd}[1]{{\color{black}{#1}}} 
\newcommand{\eat}[1]{} % This is used to sout the background section
\newcommand{\finalcut}[1]{} % will be used to replace the s o u t commands
\newcommand{\arxiv}[1]{#1}
\newcommand{\ieeeversion}[1]{}

\def\BibTeX{{\rm B\kern-.05em{\sc i\kern-.025em b}\kern-.08em
    T\kern-.1667em\lower.7ex\hbox{E}\kern-.125emX}}

\begin{document}

\title{Proposing an Interactive Audit Pipeline for Visual Privacy Research}

\author{ 
{\Large Jasmine DeHart\(^\dagger\), Chenguang Xu\(^\dagger\), Lisa Egede\(^\ddagger\), Christan Grant\(^\dagger\)}
\\$^\dagger$ School of Computer Science, University of Oklahoma \\
\{dehart.jasmine, chguxu, cgrant\}@ou.edu\\
$^\ddagger$ Human Computer Interaction Institute, Carnegie Mellon  University \\
 legede@cs.cmu.edu 
}

\maketitle

\begin{abstract}
In an ideal world, deployed machine learning models will enhance our society.
We hope that those models will provide unbiased and ethical decisions that will benefit everyone.
However, this is not always the case; issues arise \finalcut{from}\jd{during} the data \finalcut{curation}\jd{preparation} process \finalcut{to}\jd{throughout the steps leading to} the models' deployment.
The continued use of biased datasets and \jd{biased} processes will adversely damage communities and increase the cost to fix the problem \jd{later}.
In this work, we walk through the decision \jd{making} process that a researcher \finalcut{will need to}\jd{should consider} \finalcut{make }before, during, and after \jd{a system deployment}\finalcut{their project} to \finalcut{consider}\jd{understand} the broader impacts of \jd{their }research\finalcut{ and} \jd{in} the community.
Throughout this paper, we \finalcut{observe the critical decisions that are often overlooked when deploying AI, argue for the use of fairness forensics to discover bias and fairness issues in systems,} 
discuss fairness, privacy\jd{, and ownership} issues in the machine learning pipeline, 
assert the need for a responsible human-\textit{over}-the-loop \jd{methodology} to bring accountability into \finalcut{the deployed system}\jd{machine learning pipeline}, and finally, 
reflect on the need to explore research agendas that have harmful societal impacts.
We examine visual privacy research and draw lessons that can apply broadly to \finalcut{A}\jd{a}rtificial \finalcut{I}\jd{i}ntelligence.
Our goal is to provide a systematic analysis of the machine learning pipeline for visual privacy and bias issues.
With this pipeline, we hope to raise stakeholder (e.g., researchers, modelers, corporations) awareness as these issues propagate in the various machine learning phases.
% add case study

\end{abstract}

%%%%%%%%%%%%%%%%%%%%%%%%%%%%%%%%%%%%%%%%%%%%%%%%%%%%%%%%%%%%%%%%%%%%%%%%%%%%%%%%%
\section{Introduction}
% What is the issue? What question did you answer, or what problem did you solve through this?
% Why is this question or problem important?
% How have others answered this question or solved this problem in the past? Why have these approaches been tried? To what extent are the approaches adequate?
% What did you do? How are it used? How have others studied them in the past?
% What was your specific hypothesis (i.e., what did you expect to happen)?
% How did you test this hypothesis?
% What insights does your comparison reveal (i.e., why is your specific approach useful)?

%copped from nbiair
As society progresses, humans are becoming more dependent on the accessibility and convenience that technology offers.
Everyday a large amount of visual content is uploaded to Social Media Networks (SMNs)\finalcut{by smart city severs} \jd{from billions of users across the globe}, which can explain the large amounts of sensitive data \jd{that is} available online.
While these ecosystems have goals that \finalcut{center}\jd{revolve} around helping people build connections with others; there are gaps in the methods used to protect the information for individuals and corporations who share or collect \finalcut{media}\jd{content}~\cite{krishnamurthy2008characterizing, gross2005information,rosenblum2007anyone,madejski2011failure,van2016privacy,elmaghraby2014cyber,dehart2020social}.
%~\ceg{You have to add citations to support this statement.}.
The increased upload\finalcut{s} of images and videos emphasize the need for privacy protection and \finalcut{leak}mitigation strategies \jd{for visual content}.
Visual privacy techniques extend from SMNs to connected networks, smart cities, lifelogging, and much more~\cite{dehart2020considerations}. %\ceg{Introduce visual privacy research before discussing it}.
\finalcut{Different types of}\jd{Various} harms can occur as a result of sensitive information being displayed, which makes visual privacy a growing area of concern~\cite{gross2005information, li2017effectiveness,rosenblum2007anyone}. %\ceg{Add citation or discuss the harms}.,dehart2018visual,dehart2020social
Existing technologies in the industry can show a disregard for protecting the information of individuals who share visual content or for individuals who are captured in the content~\cite{madejski2011failure,elmaghraby2014cyber}.
% {These companies may argue against this, reword or cite research that supports this}.
%While research is being done to address these concerns, there exists a gap in understanding the overlap between fairness, privacy, and feedback in the Visual Privacy Machine Learning (VPML) pipeline.
While research is being done to address these concerns, there exists a gap in understanding the overlap between fairness, privacy, and user-feedback for issues \finalcut{of}\jd{regarding} visual privacy in the machine learning (ML) pipeline.

\jd{A need for} visual \finalcut{P}\jd{p}rivacy \finalcut{is a topic that }has emerged from \finalcut{concepts of Social Media Networks}\jd{SMNs and} \finalcut{to }the integration of Internet of Things (IoT) devices that \jd{can }expose sensitive information through visual \finalcut{a medium}\jd{content}~\cite{korayem2016enhancing, hoyle2015sensitive,sanchez2019smart}.
The constant sharing and storing of videos and images bring skepticism about individual privacy and rights~\cite{such2017photo, zhong2018toward}.
Researchers have created datasets, models, and deployed applications that they believe will provide privacy to its' users~\cite{Arlazarov_2019,tonge2016image,tonge2020image, li2017blur,zhong2018toward,zerr2012privacy, tierney2013cryptagram}.
Within these algorithms and systems\finalcut{ deployed}, researchers should continually make decisions to assess the fairness, privacy, \finalcut{and} accessibility of \jd{the data and model} \finalcut{of} \jd{in regard to} the communities they serve. % Nice
\cg{Bias can be curated from the data collection process, reinforced in the model's training, and systematically imposed in the deployment phase~\cite{suresh2019framework}.}
\finalcut{Several issues arise in the visual privacy
}\jd{With privacy and bias issues arising }throughout the \jd{ML}\finalcut{} pipeline it provokes the question: \emph{does the \finalcut{disproportionate }impact \finalcut{of visual privacy research}\jd{from model development procedures} outweigh the \jd{societal} benefits?}
% \ceg{I would consider emphasizing this sentence}
% 

%Concerns about bias and ethics have been seen across the board.
%Fairness and justice can be byproducts of  systems like hiring, recommending, and lending.

\arxiv{
    The goal of this paper is two-fold.
    First, the aim is to understand visual privacy and fairness \finalcut{in which}\jd{as} their issues intrude into the  pipeline and potentially impact \finalcut{on}\jd{the} stakeholders and community \finalcut{of the}\jd{where the} pipeline \jd{is deployed}.
    Second\jd{ly}, \jd{we }provide a comprehensive pipeline indicating \jd{fairness and privacy }issues and propose an auditing \finalcut{system}\jd{strategy} to \finalcut{raise awareness of the concerns from}\jd{reduce these affects in} visual privacy\finalcut{ and fairness}\jd{research}.
    In this paper, we observe the critical decisions that are often overlooked when deploying AI (\S~\ref{sec:vp-background}).
    We further discuss several privacy, fairness, and ownership issues that can arise in the  pipeline (\S~\ref{sec:define-mlp},~\ref{sec:explore-pf}).
    We argue for the use of human\finalcut{~\textit{in}}\jd{\textit{over}}-the-loop strategies to discover privacy and fairness issues in \finalcut{systems}\jd{ML}\finalcut{ and }\jd{. We }extend this technique to \finalcut{a}\jd{suggest two auditing processes:} Fairness Forensics Auditing System (FASt) and Visual Privacy (ViP) Auditor (\S~\ref{sec:audting}).
    % We also assert the need for a responsible to bring accountability into deployed system (\S~\ref{sec:hola}).
    Finally, reflect on the need to pursue research agendas that have harmful societal impacts (\S~\ref{sec:discussion}).
}

\arxiv{
\section{Background}
\label{sec:vp-background}
% describe related methods development work chronologically or topic-wise
% describe how newer methods improved on previous work, relaxed assumptions, used fundamentally different approaches, or tried to solve similar problems.
% When describing each method, say what it is and how it builds on previous approaches before describing its limitations.
% Better to teach here and tell a story than just to establish dominance in a (possibly) crowded area. Related work sections are often judged on their intellectual honesty
%
%proposed ml pipelines -- CG
%
%visual privacy systems -- JD
Visual privacy research provides technologies that can be used for multiple scenarios.
ML algorithms and privacy-aware systems are developed to mitigate individual and platform risks in the realm of visual privacy.
Algorithms and systems have been implemented for individuals in their daily lives~\cite{vonZezschwitz:2016:YCW:2858036.2858120, DarlingID, Dimiccoli2017MitigatingBP, Gurari2019VizWizPrivAD, korayem2016enhancing} and on social media networks~\cite{Zerr:2012:PSP:2396761.2398735,Li2019HideMePP, tierney2013cryptagram, Kuang2017AutomaticPP}.
To protect the visual privacy of individuals, researchers have suggested the use of concepts like obfuscation~\cite{li2017blur,padilla2015visual} for mitigating objects and faces and privacy risk scores.
%.~\cite{dehart2018visual}.
% With such a range of awareness and mitigation techniques, we explore work from the field.
The idea of visual privacy mitigation is centered around accessibility and caters to the individualistic concept of privacy~\cite{dehart2018visual, dehart2020social}.
Most visual privacy systems use one or more of these five protection techniques~\cite{padilla2015visual}: intervention~\cite{tierney2013cryptagram},
%dehart2018visual,
blind vision~\cite{tierney2013cryptagram},
%{dehart2018visual,
secure processing~\cite{zerr2012privacy, tierney2013cryptagram}, redaction~\cite{li2017blur},
%dehart2018visual, 
and data hiding~\cite{tierney2013cryptagram}.

Researchers~\cite{li2017effectiveness} have explored visual content to understand how attackers can extract textual information, including credit card numbers, social security numbers, residence, phone numbers, and other information.
The authors divide their privacy protection strategies into ones that control the recipient and others that control the content.
% The authors seek to introduce a privacy-enhancing technique for images.
% This technique uses obfuscation as a control mechanism for shared information.
They further conduct a user study to evaluate their effectiveness against human recognition and how those alterations can affect the viewing experience.
In other words, researchers~\cite{Gurari2019VizWizPrivAD} wanted to build a privacy-aware system for blind users of social media networks.
The dataset for this paper was collected via VizWiz, a mobile application that allows the participants to consent to their photos in this study.
Before making this data public, the authors removed private objects to protect each user.
% The images were annotated with object names, bounding box locations, and masking.
% This paper looks at images from a specific group of people and it has over 5,500 private images, including 23 comprehensive classes.

From these works, we notice the range of applications and the broad impact that they can have on society.
When building these algorithms and systems, issues with fairness and privacy can seep into the pipeline.
%fairness and privacy papers --  CG, JD
% Bias and visual paper
One of the most widely used models for computer vision is the Convolutional Neural Network (CNN)~\cite{hendricks2018women, simonyan2014very, ranjan2017all}.
Wang et al.~\cite{wang2020towards} studied bias on visual recognition tasks by a CNN model and provided the strategies for bias mitigation.
Measuring social biases in grounded vision and language embedding leads into a direction of studying biases from a mixture of language and vision~\cite{ross2020measuring}. 
A comparison study using multiple visual datasets is performed to help to understand how biases could be in datasets and affect object recognition task~\cite{torralba2011unbiased}.
There are significant disparities in the accuracy against darker-skinned females for the commercial gender classification systems~\cite{buolamwini2018gender}.
% Bias and visual privacy paper
Ethical concerns arise in facial processing technology.
Two design considerations and three ethical tensions for auditing facial processing technologies are described in~\cite{raji2020saving}.  
Specifically, auditing products need to be cautious about the ethical tension between privacy and representation. 
A framework for protecting users' privacy and fairness has been proposed in Sokolic et al.~\cite{sokolic2017learning}.
The framework blocks harmful tasks, such as gender classification, which can generate sensitive information to certify privacy and fairness in the face verification task.

The accuracy and precision of these systems can depend on (1) the data collection process, (2) fairness forensics performed on the data, (3) human-\textit{over}-the-loop techniques during model training, and (4) post-training evaluation.
The researchers should also consider the effects of deployed models, the trade-offs for private and fair systems, methods to mitigate harms, and additional measures that can be added for accountability.
The impact of our approach is that it addresses these existing limitations on traditional visual privacy systems and suggests auditing strategies for the ML pipeline.
This suggested pipeline considers privacy and fairness issues at each phase of the pipeline.
In an upcoming section, we explore the application of human-\textit{over}-the-loop and extend that technique to incorporate FASt and ViP Auditor to allow researchers to create safe and fair systems.%~\ceg{``We further''\ldots can you say in the next section?}.
} % End eat background

%%%%%%%%%%%%%%%%%%%%%%%%%%%%%%%%%%%%%%%%%%%%%%%%%%%%%%%%%%%%%%%%%%%%%%%%%%%%%%%%%
\section{Defining the Machine Learning Pipeline}
\label{sec:define-mlp}
\begin{figure}[ht]
  \centering
  \includegraphics[width=\linewidth]{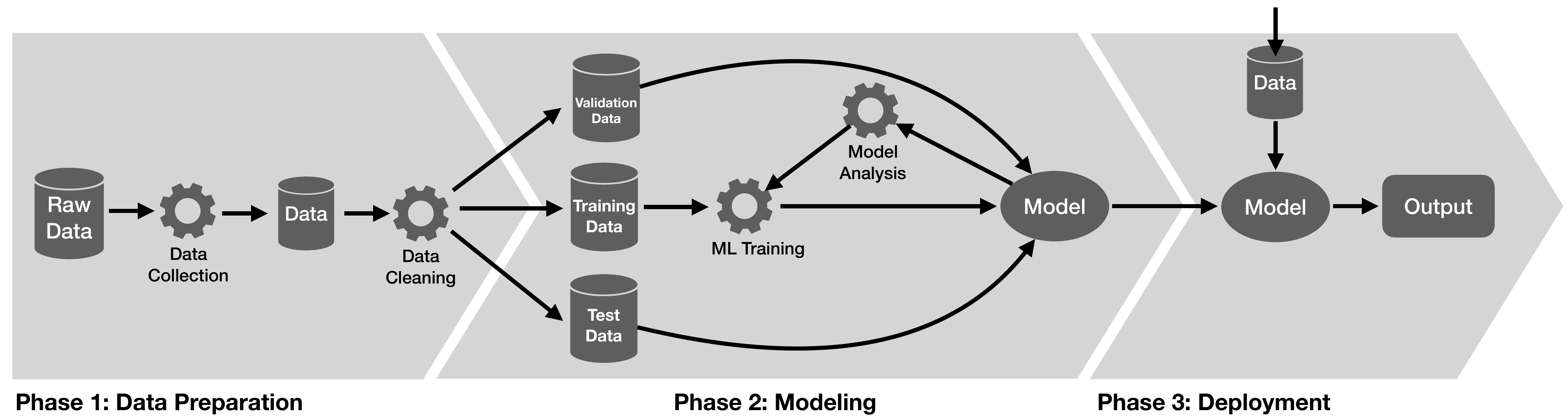}
  \caption{This figure illustrates the traditional ML pipeline\finalcut{ we described throughout the paper}. The pipeline has three phases: data preparation, modeling, and deployment.}
  \label{fig:mlp}
\end{figure}

%probably not necessary but more is better, right?
\finalcut{For this study, w}\jd{W}e describe the ML pipeline as having three phases (\Cref{fig:mlp}).
Phase 1 is the \textbf{Data Preparation} \finalcut{methodology}\jd{process}.
This phase includes considerations of (1) raw data sources, (2) data collection processes, (3) \finalcut{collected} data storage, and (4) data cleaning processes that a researcher \finalcut{w}\jd{sh}ould \finalcut{use}\jd{explore} before entering into the \finalcut{modeling}\jd{next} phase.
Data can come from anywhere and everywhere. 
With so many data source possibilities available, the researcher should consider which sources are relevant to them.
The data collection process for researchers can include using existing image datasets, social media datasets, or web scraping methods \jd{in respect to the visual privacy research task}.
% on the internet and social media. 
 Once a dataset is collected, a researcher could employ data cleaning tasks (e.g., crowd-sourced labeling) to derive an optimal dataset and labels.
%\begin{figure*}[ht]
%\end{figure*}

In Phase 2, shown in \Cref{fig:mlp}, we \finalcut{step into}\jd{begin} the \textbf{Modeling} process.
The cleaned data \jd{from Phase 1 can be}\finalcut{is} divided into \jd{three datasets: }\finalcut{a}training\finalcut{set}, testing\finalcut{set}, and validation\finalcut{set}.
Training data is used as input for the ML algorithm.
After training with the researcher's desired \finalcut{model}\jd{ML algorithm}, the researcher receives a model to run testing and validation \jd{datasets on}.
Th\jd{e}\finalcut{is} model provides the output of the \finalcut{models' }performance with several metrics.
This new information can be used for refining the model before \jd{entering the final phase of the pipeline.}\finalcut{the system is deployed.}

The last phase of the proposed machine \jd{learning} pipeline is \textbf{Deployment}.
The \textbf{Deployment} phase uses real-world data as input for the \finalcut{tuned}\jd{selected} model.
The researcher or end-user will see the real-world \finalcut{predictions}\jd{applications and results }of their selected model from Phase 2.
This phase allows the researcher to evaluate their model's performance and impact in the communities \jd{that} they serve.
%%%%%%%%%%%%%%%%%%%%%%%%%%%%%%%%%%%%%%%%%%%%%%
%%%%%%%%%%%%%%%%%%%%%%%%%%%%%%%%%%%%%%%%%%%%%%

\subsection{The Guise of Pipeline Ownership}
Researchers must consider who has ownership of the data and model at each phase \finalcut{when}\jd{before} beginning these processes. These considerations are essential when protecting the privacy of individuals and biases that the owner could impose.

At the \textbf{Data Preparation} and \textbf{Deployment} phases, the researcher should consider who are the owners of the data and how they are receiving the content.
This can explore if online images belong to the users or the corporation\jd{,} if existing datasets belong to the proprietary researchers, or \jd{if }agreements on volunteered data \finalcut{from}\jd{belong to} companies or individuals.
Furthermore, if the researcher is using online resources for data processing,\jd{ the researcher should consider:} how is the data stored\finalcut{? D}\jd{, d}oes it still belong to the researcher\finalcut{? W}\jd{, and w}hat information is being stored on these platforms\finalcut{?}\jd{.}

In the \textbf{Modeling} and \textbf{Deployment} phases, the researchers should consider who holds ownership of the model.
Considerations should be given to understand if the rights and ownership of the model are owned by the researcher, company, or from a third party~\cite{neyaz2020security}.%, stark2017faceapp}.
\finalcut{The uploaded data potentially lead to a privacy breach due to the data accessibility of the model owner.}
\jd{The data uploaded to the model could lead to an individual perceiving that their privacy has been breached due to the authorization and ownership of the model.}
If a third party owns the model, \jd{it is important to consider }what information they are collecting from the use of it\finalcut{? W}\jd{and w}ho \finalcut{do }they share this information with\finalcut{?}\jd{.}
Throughout each phase of the ML pipeline, the stakeholders should continue to ask tough questions and make critical decisions that are ethical, fair, and in the best interest of those \finalcut{impacted}\jd{the technology is meant to serve}.

%%%%%%%%%%%%%%%%%%%%%%%%%%%%%%%%%%%%%%%%%%%%%%%%%%%%%%%%%%%%%%%%%%%%%%%%%%%%%%%%%

\section{Exploring Privacy and Fairness Concerns in the VP Machine Learning Visual Privacy ML Pipeline}
\label{sec:explore-pf}
% \subsection{Potential Visual Privacy Pipeline Issues}
%- add definition table, edit errors
%- discuss major issues with developing technology and deploying
% at which stage do these issues arise
% common mitigation strategies table? research paper, understood issues, mitigation strategy
Efforts to implement technology that serves to mitigate harm are the motivation behind visual privacy research.\finalcut{Positive outcomes are desired from \finalcut{software}\jd{systems} that \finalcut{is}\jd{are} created to help solve society's most pressing issues, such as privacy leakage on social media.}
In this section, we will discuss privacy \finalcut{concerns }and fairness issues that frequently occur with developing and deploying visual privacy systems\finalcut{,}\jd{. Examples of these issues are} shown in~\Cref{fig:mlp_issues} and \Cref{table:tab_issue}.
We suggest that \finalcut{mitigating}\jd{when evaluating} visual privacy \finalcut{issues}\jd{systems, researchers} should consider bias \jd{issues }as \finalcut{it}\jd{they} arise\finalcut{s} in the ML pipeline.
\arxiv{
As apparent from Figure~\Cref{fig:mlp_issues} there are fairness issues involved with all stages of ML pipeline. 
This investigation will compromise of three over-arching visual privacy issues and describe how they could affect the ML pipeline.
The question becomes are these models necessary, and do they serve to benefit the people or system that is affected by it?
% Each issue will be defined and will discuss at which stage do these issues arise
% In this work, we focus on fairness issues that frequently occur in a machine learning pipeline and could have influence on visual privacy issues.
% We describe several fairness issues as shown in Table~\ref{table:tab_issue}.
}
% In this section, we will describe the typical fairness issues.

\begin{table*}
%\begin{table}
  \caption{This table displays the privacy and fairness issues in various phases of the machine learning pipeline.
  The description provides a high-level overview of what those issues are.
  The checkmark (\checkmark) indicates that those issues could arise in those parts of the pipeline.
  }
\centering
\begin{tabular}{ l|p{1in}|p{2.4in}|p{0.1in}:p{0.1in}:p{0.1in}:p{0.1in}|p{0.1in}:p{0.1in}:p{0.1in}:p{0.1in}|p{0.1in}:p{0.1in}:p{0.1in}| } 
%\multicolumn{1}{c}{} &\multicolumn{1}{c}{} & & \multicolumn{2}{c|}{Data} &\multicolumn{3}{c|}{ }&\multicolumn{2}{c|}{ } \\
\multicolumn{1}{c}{} & \multicolumn{1}{c}{}& & \multicolumn{4}{c|}{Phase 1} & \multicolumn{4}{c|}{Phase 2}& \multicolumn{3}{c|}{Phase 3}\\
\cline{4-14}
\multicolumn{1}{c}{}& \multicolumn{1}{c}{}&  Description & \rot{Raw Data} & \rot{Data Collection }& \rot{Data} & \rot{Data Cleaning} & \rot{Training Data} & \rot{ML Training}& \rot{Model}  & \rot{Model Analysis} &\rot{Data} &\rot{Model}& \rot{Output} \\ 
\cline{2-14}
 & Obtaining Content Consent  &  Exploring the ethics and privacy methodology of researchers to obtain consent for collected visual content on a public domain.  &  &\checkmark & & \checkmark  & & &  & & \checkmark &  & \checkmark \\ 
%\cdashline{2-14}
\rowcolor[gray]{.9}
\cellcolor{white} & Multiparty Conflict & Understanding privacy concerns for images and videos which are owned by multiple persons.
  &  \checkmark &  \checkmark  & \checkmark  & \checkmark & &\checkmark & & \checkmark  & \checkmark &  & \checkmark \\ 
%  \cdashline{2-14}
\multirow{-5}{*}{\rot{Privacy issues}}
&Image Removal Request & Determining when and how visual consent should be removed from the pipeline via requests.  & &  & \checkmark & \checkmark & & \checkmark  &   \checkmark & \checkmark  & \checkmark  &   & \checkmark \\ 
\hline
\rowcolor[gray]{.9}
\cellcolor{white} 
& Historical bias & The inherent bias from a biased world is absorbed by the source data.~\cite{suresh2019framework}  & \checkmark & & & &  & & & &\checkmark & &\\ 
%\cdashline{2-14}
& Algorithmic bias& The bias relates to the algorithm in the ML pipeline, and it could have different bias sources and types.~\cite{bantilan2018themis}~\cite{calmon2017optimized}~\cite{danks2017algorithmic} & &  & & & \checkmark &\checkmark & \checkmark & \checkmark  & & \checkmark & \checkmark\\ 
%\cdashline{2-14}
\rowcolor[gray]{.9}
\cellcolor{white} &Software \newline Discrimination& The output from a predictive software used to aid in decision making may lead to unfair consequences. ~\cite{galhotra2017fairness} & &  & & & & & & & & \checkmark & \checkmark \\ 
%\cdashline{2-14}
&Individual fairness& Similar individuals should be treated as similarly as possible.~\cite{dwork2012fairness} & & & & &\checkmark &\checkmark &\checkmark &\checkmark & & \checkmark& \checkmark\\ 
%\cdashline{2-14}
\rowcolor[gray]{.9}
\cellcolor{white} &Group fairness& The groups defined by protected attributes should obtain similar treatments or equal opportunity as the privileged group.~\cite{hardt2016equality} & & & & &\checkmark & \checkmark&\checkmark & \checkmark & & \checkmark& \checkmark\\ 
%\cdashline{2-14}
&Disparate treatment& Protected attributes are directed applied in the process of modeling where unfairness occurs.~\cite{zafar2017fairness}&  && & & \checkmark& \checkmark & \checkmark&\checkmark & &\checkmark &\checkmark \\ 
%\cdashline{2-14}
\rowcolor[gray]{.9}
\cellcolor{white} \multirow{-15}{*}{\rot{Fairness issues}}  &Disparate impact& Even though the protected feature is not directly using, its relevant features still could lead a selection process to make unfair outputs. ~\cite{feldman2015certifying} &&  & & & \checkmark&\checkmark&\checkmark &\checkmark & &\checkmark &\checkmark \\ 

%\cdashline{2-11}
%& ? Unwarranted \newline Associations & ~\cite{tramer2017fairtest}  & & \checkmark & & & & & & \\ 
\cline{2-14}
  \end{tabular}

      \label{table:tab_issue}
%\end{table}
\end{table*}

\subsection{Privacy}
\label{sec:privacy}
Visual privacy issues can arise at any point in the ML pipeline.
The stakeholders must be aware of these issues and develop ways to solve \finalcut{these issues}\jd{them proactively as }\finalcut{before or after }they arise.
\subsubsection{Obtaining Visual Content Consent}
Researchers can use large public image \jd{data} sets~\cite{zerr2012privacy, lin2014microsoft, imagenet_cvpr09} to train \finalcut{visual privacy models}\jd{ML algorithms} to perform various \jd{visual privacy research} tasks~\cite{tonge2016image,tonge2020image,Zerr:2012:PSP:2396761.2398735}.
Additionally, when collecting a large amount of data, many researchers question the use of \textit{web scraping} methods to obtain this data~\cite{zimmer2010but, zimmer2017internet, mancosu2020you,krotov2018legality}.
Large data sets can be labeled using \finalcut{teams of crowd workers and }crowd-sourcing methods~\cite{lin2014microsoft, imagenet_cvpr09, 5539970, 4531741}.
\finalcut{In place of }\jd{While researchers' efforts can focus on }creating systems \finalcut{devoted to}\jd{to help with visual} privacy, their approach \finalcut{to}\jd{in} collecting data can bring rise to privacy and ethical concerns in Phase 1 of the \jd{macine learning }pipeline.
The method\jd{s}\finalcut{ology} \cg{that} researchers use to collect this data can overlook individuals' privacy, consent, and protection.
When collecting visual content or using existing datasets, researchers can un-intentionally collect private content containing minors\finalcut{,} \jd{or} bystanders~\cite{7945228,Dimiccoli2017MitigatingBP,Hasan2020AutomaticallyDB,DarlingID,birhane2021multimodal}\jd{.}\finalcut{ or visual content not meant to be used in application in the deployed systems.}
The topic of consent is essential to gauge participants' willingness to participate in the study or research.
For traditional studies that \finalcut{use}\jd{include} people \jd{or living subject}\cg{s}, \finalcut{some}specific procedures and policies need to be followed \jd{according to a governing entity} (i.e., institutional review board), so what excludes visual privacy research from policies and procedures when using \finalcut{individuals}\jd{personal} data?

\subsubsection{Multiparty Conflict (MPC)}
Images and videos can be owned by multiple people~\cite{10.1145/3025453.3025668}.
Co-ownership \jd{issues} can arise from several situations; a few are (1) individuals engaging in group photos, (2) a person \finalcut{holding the}responsib\finalcut{ility of}\cg{le for} other\finalcut{s} \cg{individuals} (e.g., children, pets), (3) a person having physical possession of images \finalcut{with}\cg{of} others on their device~\cite{zemmels2015sharing}.
These types of conflict can affect the privacy of minors~\cite{lwin2008protecting, batool2020exploring} and bystanders~\cite{DarlingID, Li2019HideMePP, 7945228}.
This co-owned content can cause privacy leaks \finalcut{on social media} without it being the \finalcut{poster's}\cg{individual's} intent~\cite{dehart2018visual}.
In the ML pipeline, the researcher should consider possible issues for MPC in all phases.

Considerations for \cg{content} ownership and \cg{individual} rights should be made early in the pipeline.
When working with visual content, it can be necessary to seek permission from all parties involved.
Multiparty Conflicts can enter the ML pipeline as early as the \textbf{Data Preparation} phase.
In the \textbf{Deployment} phase, the \finalcut{visual content}\cg{real world data} used for the \finalcut{prediction} \cg{ML task} can bring rise to this issue.
% 
% 
% On Social Media Networks, photos can be owned by multiple people.
% \subsection{Phase 1: Data Preparation}
% \begin{enumerate}
%   
%   \item Issue: Multiparty Conflict
%     \item Issue: Demographic Classifiers
% \end{enumerate}

\subsubsection{Image Removal Requests}
When collecting data and using existing datasets, ownership issues will arise and should be addressed early and appropriately. Instead of using ``public'' resources, \cg{researchers should seek} participation consent from individuals\cg{. This} becomes important when using \finalcut{this}\cg{data} for research and \cg{in} deployed systems. This \finalcut{begs}\cg{raises} the question, what to do if an individual's visual content is requested to be removed from the dataset and the model's training \cg{phase}? In July 2020, MIT decided to remove the 80 Million Tiny Images dataset because of the bias and offensive labels that occurred in the dataset~\cite{torralba_fergus_freeman_2020}. If researchers have used this dataset, these issues can affect the credibility of the\jd{ir} work and the deployed system\jd{,} if one exists.
Image removal requests can affect all phases of the ML pipeline \cg{and should be handled accordingly}.
% removed severely
%%%%%%%%%%%%%%%%%%%%%%%%%%%%%%%%%%%%%%%%%%%%%%%%%%%%%%%%%%%%%%%%%%%%%%%%%%%%%%%%%
% 
% \section{Exploring Fairness Issues in the Machine Learning Pipeline}
% In this work, we focus on fairness issues that frequently occur in a machine learning pipeline and could have influence on visual privacy issues.
% We suggest that mitigating visual privacy issues should consider the bias that can arise in the machine learning pipeline.
% As apparent from Figure~\ref{fig:mlp_issues} there are fairness issues involved with all stages of ML pipeline. 
% %We describe several fairness issues as shown in Table~\ref{table:tab_issue}.
% In this section, we will describe the typical fairness issues.
\subsection{Fairness}
% Explain why 7 issue goes to 3 
In this section, we\finalcut{talk about the} discuss three typical fairness issues.
These biases sneak into most steps of the ML pipeline and could propagate to other parts of the pipeline. 
These three general issues can lead researchers to think about where or when \finalcut{the fairness issue} \cg{bias can} occur\finalcut{s}.
Later, in the algorithmic bias \cg{section}, we will discuss four more specific biases (i.e., \textit{individual fairness} versus \textit{group fairness}, \textit{disparate treatment} versus \textit{disparate impact}) that explore\finalcut{s} who is affected and how those issues arise in the pipeline.
% Those four specific issues take care of who is affected and how those issues arise.

\subsubsection{Historical Bias}
When data is generated, the inherent bias from the world could stealthily engrave into data.
Historical bias can enter the ML pipeline at the start point of the \textbf{Data Preparation} phase and the \textbf{Deployment} phase.
Even under ideal sampling and feature selection, historical bias could still exist and cause concern.
When the historical bias proliferates through the ML pipeline, it can impact modeling and decision-making in the deployment stage~\cite{hellstrom2020bias, suresh2019framework}.
  
\subsubsection{Algorithmic Bias}
Algorithmic biases are bound together with each process in the ML pipeline.
Roughly, algorithmic bias is focused in the \textbf{Modeling} phase.
Because algorithms are connected with every part of ML systems, there are different bias sources and types from different components of the ML pipeline. 
The algorithm's bias could be sourced from biased training data, a biased algorithm, or misinterpretation of the algorithm's output~\cite{danks2017algorithmic}. 
Identifying the source of algorithmic bias contributes significantly to dissolving the fairness issue.
In addition, we must also consider the types of algorithmic bias.
%Group versus individual fairness
Usually, we can start \cg{to} think\finalcut{ing} about who is the victim impacted by the algorithmic bias. 
For example, similar individuals are treated inconsistently based on the predictions of the model, while \textit{individual fairness} requires that each similar individual should be treated as similarly as possible~\cite{dwork2012fairness}.
As a more general example, \textit{group fairness} considers groups defined by protected attributes (e.g., gender, race), and it requires that the protected groups should obtain similar treatments as the privileged group~\cite{hardt2016equality}.
\textit{Group fairness} is also referred to as statistical parity or demographic parity.

%Disparate treatment versus disparate impact
After identifying who suffers from the algorithmic bias, it becomes increasingly important to understand how fairness issues arise in the ML pipeline.   
\textit{Disparate treatment}, also known as direct discrimination or intentional discrimination, occurs when protected attributes are used explicitly in the ML system. Consequently, disadvantaged groups identified by the protected attributes are deliberately treated differently.
\textit{Disparate impact} referred to indirect discrimination or unintentional discrimination, is pervasive and entrenched in our society~\cite{feldman2015certifying}.
Regarding \textit{disparate impact} in the ML pipeline, it exists under the guise of correlated variables that implicitly correspond to the protected attributes.

\subsubsection{Software Discrimination}
Last but not least, software discrimination appears at the end of the entire ML pipeline, which is the \textbf{Deployment} phase, and bias could still exist due to the problematic model.
After an ML model is passed to its end-users, the interpretability and transparency of the model can benefit from identifying and mitigating potential bias generated by the software.
Researchers have developed many tools that audit fairness for a deployed ML model.
Tools like IBM's AI Fairness 360 toolkit~\cite{bellamy2019ai} implement fairness metrics and bias mitigation algorithms.
Other works have generated test suites to measure software fairness from a causality-based perspective~\cite{galhotra2017fairness}.

\subsection{Overlaps in Privacy and Fairness Issues}
%cases in which both arise
%trade-offs
%indicate pipeline overlaps with colors
%how can one lead to another
%\begin{figure*}[ht]
\begin{figure}[ht]
  \centering
  \includegraphics[width=\linewidth]{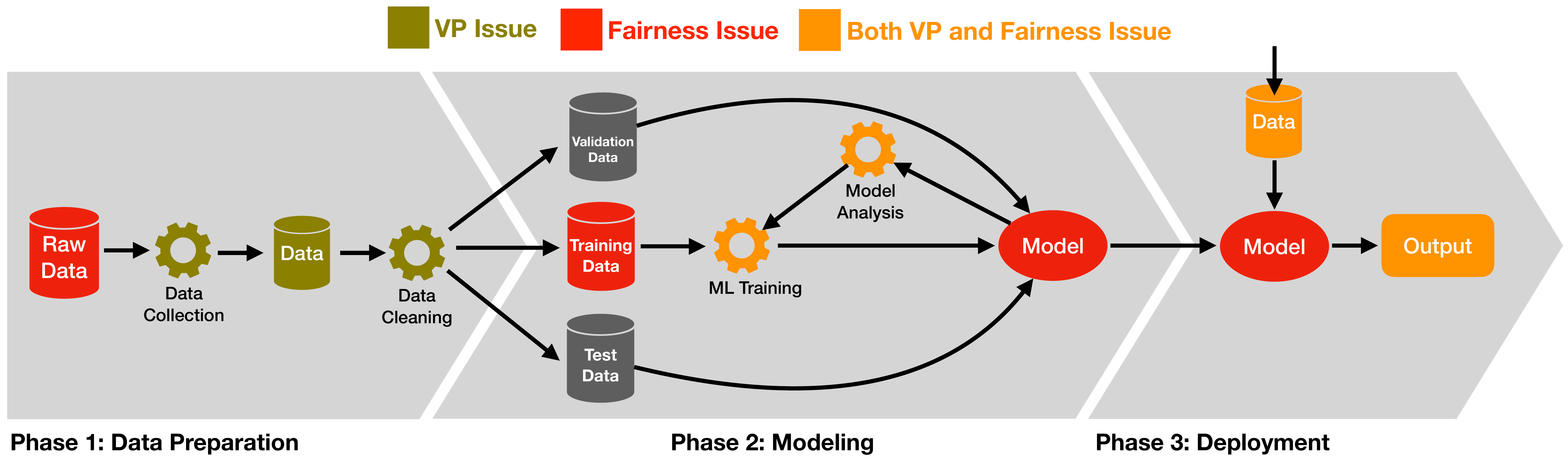}
  \caption{In the ML pipeline, we indicate where privacy (green) and fairness (red) issues could arise.
  Possible overlaps in the system are defined in orange.}
    \label{fig:mlp_issues}
\end{figure}
%\end{figure*}
From \Cref{fig:mlp_issues}, we can observe that some steps that contain both visual privacy and fairness issues\finalcut{ could occur}. 
\finalcut{In addition, some issues impact the point where the issue occurs and raise concerns in the later phases.}When both issues arise, \finalcut{we have}\cg{researchers should}\finalcut{to} be ready to deal with them; otherwise, they will affect the system's outcome. 
For instance, a model builder perceives that the protected groups could be affected by the fairness issues in a facial recognition system.
Consequently, the modeler strives to collect more data to make up for the disproportion. 
However, the privacy risk for the data collection could be an unexpected problem and is increased for the underrepresented group~\cite{raji2020saving}.

\arxiv{
It is essential to understand the relationship between visual privacy issues and fairness issues, since sometimes solving one type of issue could have a negative impact on the other type of issues.
For instance, a user uploaded a picture to a biased ML model in the cloud.
The user could experience unfair decisions from the biased model with simultaneous loss of privacy to the service provider.
One of our goals is to raise awareness of such worse cases.
Trade-off analysis between privacy and fairness will develop an in-depth understanding of the building process for such a complex system.
}
% 
%%%%%%%%%%%%%%%%%%%%%%%%%%%%%%%%%%%%%%%%%%%%%%%%%%%%%%%%%%%%%%%%%%%%%%%%%%%%%%%%
\section{Integration of Interactive Audit Strategies for the Machine Learning Pipeline}
\label{sec:audting}
% 
%Once you deploy a machine learning model is it possible to reel it back in.
ML models are constantly being updated once deployed to the real world;
\finalcut{Frequent}\cg{regular} updates help to avoid and minimize costly errors.
Differences in the time between error discovery and model correction for the deployed model are crucial.
%At what point do we decide to enter the VP/ML Pipeline? Some biases can't be mitigated \& I think that would be important to define before entering that pipeline
%What are the Audit Phases?
% A comprehensive visual privacy system can include (1) data collection processes, (2) a trained model, (3) a feedback system, and (4) a user interface.
Systems should be able to respond to unexpected bias before, during, and after deployment.
%ystem can include (1) data collection processes, (2) a trained model, (3) a feedback system, and (4) a user interface.
%
%elaborate on points for this section need tie into the pipeline sections
% The importance of the collection and training processes becomes critical as these models are deployed in SMNs and IoT devices.
\finalcut{Adding human decision-making in the ML systems with a human-\textit{in}-the-loop framework introduces supervision %oversight
into deployed models.}
% To actively

% \subsection{Human-\textit{in} or \textit{over}-the-loop?}
It could be impossible to erase the damage caused by the aftermath of a system; however, stakeholders could start making a change now.
One way to do this would be using an interactive ML  approach, human-\textit{in}-the-loop
% or Automated Machine Learning (AutoML) 
~\cite{fails2003interactive,amershi2014power, amershi2015modeltracker, DBLP:journals/debu/LeeMXLHP19}.
%A human-\textit{in}-the-loop approach can resolve learning issues that arise in the pipeline when the models' training is not supervised.
Training in the human-\textit{in}-the-loop framework requires humans to make incremental updates to anticipate issues~\cite{bond2016framework}.
Traditional ML pipelines conduct training on their own without interference from humans.
To debug these models, the researcher must begin a thorough investigation of the models' predictions, parameters, and data after the learning has been completed.
An interactive approach would allow a person in the model's training process, which will reduce debugging and \finalcut{run time}\cg{runtime}.
The \jd{human}\finalcut{person, in this case, referred to as the human,} is able to check the learning for the model and coach the model to meet the desired result in a feedback cycle.
\finalcut{The f}\jd{F}eedback cycles allow the researcher to provide positive feedback iteratively to the model after viewing the processes.
This can allow the \finalcut{modeler}\jd{researcher} to understand the possible bias and privacy issues in the model and mitigate it immediately.
This approach can be extended to various ML research areas in fairness, computer vision, and privacy.

In traditional human-\textit{in}-the-loop approaches, the human becomes a bottleneck for the feedback process.
In light of this, we suggest using a \textbf{human-\textit{over}-the-loop }approach\finalcut{ described in}~\cite{graham2017formalizing}.
Human-\textit{over}-the-loop allows researchers to step into the pipeline as needed to perform corrections.
This removes the \finalcut{hassle}\jd{necessity} of a human approving each iteration of the model.
With this feature integrated in the ML pipeline, the researchers should consider having multiple ``humans'' to monitor the training.
This, in turn, can\jd{ lower response times to } resolve biases that may be imposed from ``humans'' during learning.
%Having additional ``humans'' introduces a check in the system to ensure one person does not have full power over training.
% If issues arise during the deployment, the human- \textit{over}-the-loop could become the fall guy.
Based on the human-\textit{over}-the-loop technique, we propose the use of two interactive auditing strategies that can reduce fairness and privacy issues to allow researchers to conceptualize, develop, and deploy safer visual privacy systems.
%\begin{figure*}[ht]
\begin{figure}[ht]
  \centering
  \includegraphics[width=\linewidth]{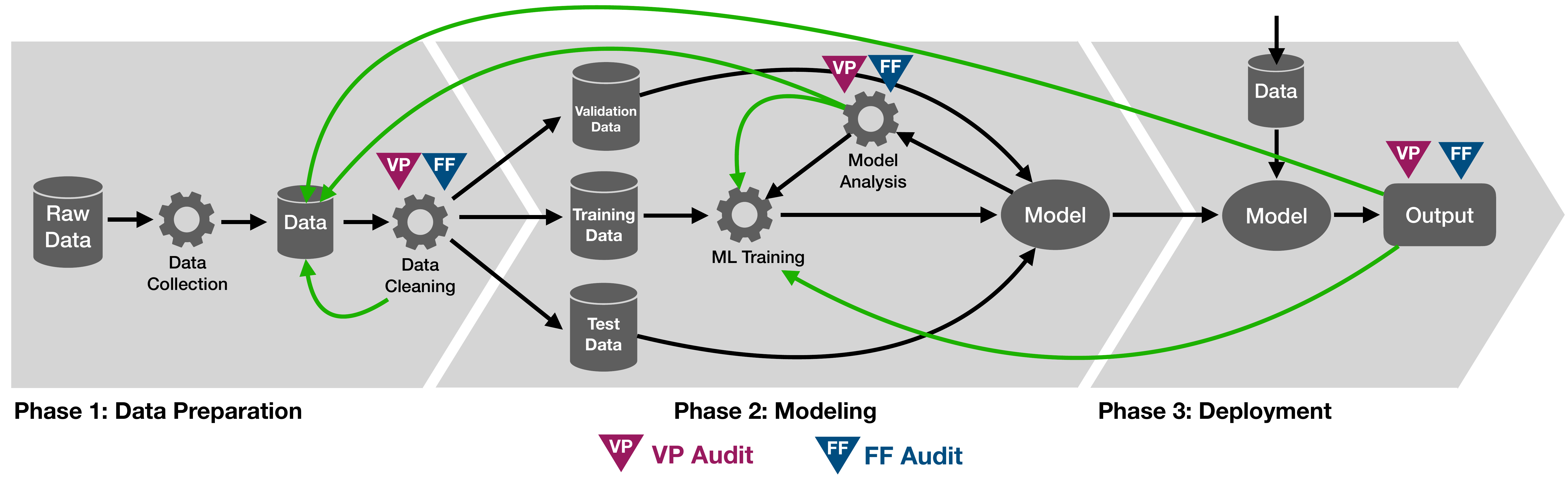}
  \caption{This figure illustrates the feedback loops when human-\textit{over}-the-loop techniques are implemented.
  The green lines denote audit traces for feedback loops.
  The loops that are suggested to have a ViP Audit are denoted with a VP marker.
  The loops that are suggested to have  FASt are denoted with a FF marker.
  Data Preparation has one feedback loop from Data to Data Cleaning.
  The Modeling phase has two feedback loops: (1) from model analysis to ML training, and (2) from Model Analysis to Data (in the Data Preparation phase).
  In the final phase of the pipeline, the deployment loops are from (1) Output to Data (in the Data Preparation phase) and (2) Output to ML training in the Modeling phase.
%   We suggest the use of FASt and ViP to enhance feedback in the areas of fairness and privacy.
  }
    \label{fig:mlp_feedback}
\end{figure}
%\end{figure*}
%%%%%%%%%%%%%%%%%%%%%%%%%%%%%%%%%%%%%%%%%%%%%%%%%%%%%%%%%%%%%%%%%%%%%%%%%%%%%%%%%
\subsection{Fairness Forensics Auditing System (FASt)}
% COPY FROM Navigating the Broader Impacts of AI Research
% Need to modify
% Consider feedback loops at each stage
ML bias is a rising threat to justice, and it has been investigated in broad areas, including employee recruitment, criminal justice, and facial detection.
ML research can cause unanticipated and harmful consequences on our daily life while decision-makers begin to utilize the result of the output from ML algorithms without considering fairness.
Fairness forensics focuses on supporting data scientists and modelers to inspect a dataset or a new model by techniques and tools evaluating for bias.

Fairness forensics requires overarching understandings of the types of bias, the entire pipeline of ML systems, and the analysis of bias on different pipeline stages.
It is vital to understand how biases have harmful impacts on different communities of people when deploying ML systems related to visual privacy in societal domains.
Fairness forensics has three major tasks: bias detection, bias interpretation, and bias mitigation. 
People can use fairness metrics to evaluate the input or output of ML models for bias detection.  
Bias report generator tools and bias visualization tools facilitate analysis and interpretation of bias for humans to understand the meaning and impact of bias detection results.
Once the bias is discovered, bias mitigation strategies can be applied by the interventions to the input data, the algorithm, or the predictions.
Bias mitigation algorithms can be categorized into three types: pre-processing, in-processing, and post-processing algorithms~\cite{bellamy2019ai}.

%add the figure and talk abt what its describing
%%%%%%%%%%%%%%%%%%%%%%%%%%%%%%%%%%%%%%%%%%%%%%%%%%%%%%%%%%%%%%%%%%%%%%%%%%%%%%%%%

\subsection{Visual Privacy (ViP) Auditor}
%formally known as ViPA
%what are privacy leaks? how can this reduce them?
%how can this reduce primary vp issues in systems?
For actively investigating visual privacy research, we propose the use of a human-\textit{over}-the-loop technique specifically designed to handle privacy, classification, and computer vision issues.
Visual privacy systems \finalcut{can}are comprise\cg{d of} multiple \finalcut{ml}\cg{ML} techniques and strategies.
We envision the ViP Auditor as a comprehensive auditing tool that will enable researchers to use visual analytics~\cite{liu2017towards} to understand the models' learning process.
During learning, the modeler will be able to enhance the feedback process by using similar schemes as ModelTracker~\cite{amershi2015modeltracker} or Crayons Classifier~\cite{fails2003design}.
The modeler will protect individual privacy in the learning process by incorporating visual privacy mitigation strategies built into the ViP Auditor.
For \finalcut{Model Analysis}\cg{Model Analysis}, the researcher can understand the dataset attributes (e.g., number of faces, number of privacy leaks for each category), the models' classification performance, and the perceived privacy risk score of the model.

% From the feedback loops in Figure~\ref{fig:mlp_feedback}, the loops that are suggested to have a ViP Audit are denoted with a VP marker.
\subsection{Understanding Pipeline Integration}
From the \textbf{Data Preparation} phase in \Cref{fig:mlp_feedback}, researchers examine the dataset, data labels, and ownership for the content regarding privacy concerns.
This loop allows a researcher to consider the initial privacy concerns (in \Cref{sec:explore-pf}) and develop other strategies to mitigate them.
% The Modeling phase has two feedback loops: (1) from model analysis to ML Training and (2) from Model Analysis to Data (in the Data Preparation phase).
In the \textbf{Modeling} \jd{p}hase, the researcher should employ auditors at both feedback loops (see \Cref{fig:mlp_feedback}).
The first feedback loop allows the researcher to conduct a privacy evaluation from the model's output.
Evaluating the results from this feedback loop enables the human-\textit{over}-the-loop to step in and make changes to achieve the desired level of privacy in the model.
Auditing at this phase of the pipeline allows researchers to accurately correct recognition errors (bounding boxes, instance segmentation) from the models' learning.
%enhance recognition/bounding boxes
The second feedback loop conducts a privacy evaluation that allows the researcher to identify issues within the dataset from the Model Analysis.
When the dataset issues are identified, the researcher can collect more data, remove the data from the pipeline, or add more tags/labels to mitigate privacy concerns that arise.
% In the final phase of the pipeline, the deployment loops are from (1) Output to Data (in the Data Preparation phase) and (2) Output to ML Training (in the Modeling phase).
The \textbf{Deployment} \cg{p}hase feedback loops consider the real-world output from the model.
With auditors in place at this phase, the stakeholders can understand privacy issues as they arise.
The stakeholders can fix issues in deployment as they arise by sanitizing the data and re-training the model.
The ViP Auditor will produce a privacy risk score based on the models' performance and flag potential privacy issues.

The feedback loop from \cg{FASt} is similar to the loop from the ViP Auditor. 
\cg{F}airness forensics \jd{system} feedback occurs at different steps in all three phases 
\jd{of the ML pipeline} (see~\Cref{fig:mlp_feedback}), and it can encourage researchers to sanitize their data or adjust the model.
The process of fairness forensics allows the human-\textit{over}-the-loop to determine the need for human intervention and assess for fairness in order to achieve social justice.
\finalcut{When ML pipeline needs to take fairness into account, humans should realize \finalcut{and balance the} trade-offs among performance metrics (e.g., accuracy), privacy, and equity.}\cg{Imperfect fairness metrics or conflicting fairness objectives~\cite{friedler2021possibility} means humans will need to intervene to maintain performance guarantees.} 

% When and how to integrate FF/HiL
% bias and privacy mitigation at each step
% \subsubsection{Integrating Human-\textit{in}-the-Loop}
%What is it? How does it help VP? Pros/Cons? Integration? How will this help with VP?
% Human over-the-loop is required.
%     - The human over the loop is the "fall guy".
%     - There may be and should be more than one human because that human is biased. (This human needs his power checked)

% \subsubsection{Enhancing the ML Pipeline with Fairness Forensics}
%What is it? How does it help VP? Pros/Cons? Integration? How will this help with VP?
%Will FF be in pre-processing and post? How this fit in with the traditional ML Pipeline?
%Fairness

%%%%%%%%%%%%%%%%%%%%%%%%%%%%%%%%%%%%%%%%%%%%%%%%%%%%%%%%%%%%%%%%%%%%%%%%%%%%%%%%%
% \section{Case Study}
% % from practice to real world when adding these methods
% % could use a related paper or VP issue for use case
% Put on hold in lieu of Section 4.3, if needed will add in for the final draft.

%%%%%%%%%%%%%%%%%%%%%%%%%%%%%%%%%%%%%%%%%%%%%%%%%%%%%%%%%%%%%%%%%%%%%%%%%%%%%%%%%
\section{Discussion}
\label{sec:discussion}
%taken from nbair
Being mindful of the societal impacts, \finalcut{pre-deployment and monitoring the ML data quality have been presented as mitigation techniques.}\jd{evaluation methods (i.e., FASt and ViP) and monitoring strategies (i.e., human-\textit{over}-the-loop) have been presented as mitigation techniques to reduce errors in the ML pipeline and in the deployed system's life cycle.}
However, there are no \finalcut{foolproof}\jd{full-proof} techniques for ensuring that the software is exempt from producing harm.
\finalcut{Knowing}\jd{For a stakeholder to know} when to halt deployment \finalcut{means having}\jd{implies that they have developed} a \finalcut{ongoing}plan for the \finalcut{product}\jd{system} and \finalcut{viewing it as a system} requir\finalcut{ing}\jd{e} \finalcut{frequent surveillance and updates}\jd{human intervention throughout the ML pipeline for proactive decision making}.
Monitoring for \finalcut{data-related}\jd{privacy and fairness} issues \finalcut{or}\jd{and} their potential to harm\finalcut{s} \finalcut{users}\jd{the community} throughout the software's life is a\jd{n essential} part of this.
\finalcut{For fairness forensics}\jd{When evaluating the fairness of a model}, \finalcut{this may look like measuring a training model's error rate and fairness metrics and deciding what results are considered successful. }\jd{a researcher can explore the model's training data and performance metrics to decipher sub-trends and anomalies.
From this evaluation, the researcher can generate an idea of what success can look like from their model.}

It might also be helpful to pivot directions for the machine learning model to avoid going too far down a path that could prove disastrous for marginalized communities.
There may be a point at which the model \finalcut{has been altered}\jd{is} beyond \finalcut{the point of} recognition.
It may be worth completely re-imagining the \finalcut{product}\jd{ML pipeline} or abandoning the effort altogether when it has strayed far from its' intended goal.
Before completely re-imaging or abandoning the model, the researcher could integrate human\jd{-\textit{over}}\finalcut{over}-the-loop techniques to improve \jd{the ML pipeline's consideration for privacy and fairness.}\finalcut{performance and fairness.}
Th\finalcut{is}\jd{e} decision of which route to go ultimately involves the researcher\finalcut{'s value of }\jd{evaluating the trade-off between the }safety for \jd{the} impacted \finalcut{individuals}\jd{communities or the} \finalcut{over the}potential accomplishments of producing innovative software.
Success should be inspired by \finalcut{ideas'}\jd{the} ability to impact society positively, not by a \finalcut{product's}\jd{system's} ability to \jd{quickly} solve a\jd{n} \finalcut{paper's }idea.
Questioning when to stop \jd{in the ML pipeline} should \jd{include} prioritizing\finalcut{how} socie\finalcut{ty}\jd{tial} \finalcut{is}impact\finalcut{ed} and \finalcut{how }the \jd{affect on} marginalized communities\finalcut{ are affected}.

Halting deployment on a project that has gone awry should be seen as a successful learning result, not a\jd{s a} fail\finalcut{ure}\jd{ed project}.
If permissible, \jd{the stakeholder should consider }opening up the research \jd{project} or \finalcut{product}\jd{system} for external \finalcut{inspection can}\jd{review to} \finalcut{generate}\jd{cultivate} a \finalcut{more}meaningful conversation around learning from the harms that \jd{development and }deployment could have caused.

%%%%%%%%%%%%%%%%%%%%%%%%%%%%%%%%%%%%%%%%%%%%%%%%%%%%%%%%%%%%%%%%%%%%%%%%%%%%%%%%%
\section{Conclusion}
\label{sec:conclusion}
%taken from nbair
Researchers should closely monitor \finalcut{the}data \finalcut{collection}\jd{preparation}\finalcut{process}, \finalcut{the }model\jd{ing}\finalcut{'s training process}, and\finalcut{ even the} deployment \finalcut{results}\jd{processes} to avoid harming \finalcut{individuals}\jd{communities} and stakeholders.
The decision making process for researchers can be \finalcut{a }challenging\finalcut{ task}, but it is imperative to continually evaluate to improve the model's learning \jd{process }and \jd{the deployment }outcomes for \jd{the} communities \jd{they serve}.
% The data used to train models should represent all subjects under a given domain, including various races, gender, age groups, etc.
% Along with monitoring the data collection process, it should be a priority to ensure that the data curation process is void of labeling language that is prejudice or discriminatory~\cite{prabhu2020large}.
When building a visual privacy model needing large amounts of data, it can be easy to obtain datasets that are \finalcut{already }widely distributed, but \jd{may not have}\finalcut{have not} been examined for discriminatory, priva\jd{te}\finalcut{cy issues}. or \jd{fairness issues}\finalcut{biased data}.
This work discusses privacy and fairness issues that frequently occur in \finalcut{a visual privacy }\jd{the }ML pipeline that could \finalcut{influence }\jd{emerge at} various phases.
We also assert the need for \finalcut{a }responsible auditing system\jd{s} to bring accountability into model training and the deployed system.
To do this, we propose using human-\textit{over}-the-loop strategies to introduce interactive auditing for fairness \finalcut{forensics} and \finalcut{visual}privacy.
With \finalcut{more fairness, privacy,}\jd{ML pipeline audits} and engaged researchers, \finalcut{the healthier and more stable visual privacy research will be.}\jd{the evaluation and consideration given to project development and deployed systems can become a standard procedure.}
Th\finalcut{is}\jd{ese} proposed \jd{mitigation strategies}\finalcut{system is} \jd{are }\finalcut{only one part }\jd{the first steps} of a much needed effort to \jd{address} privacy and fairness issues in the ML pipeline\finalcut{, but it is an important part that can make a big difference}.

%%%%%%%%%%%%%%%%%%%%%%%%%%%%%%%%%%%%%%%%%%%%%%%%%%%%%%%%%%%%%%%%%%%%%%%%%%%%%%%%%
\section{Future Work}
\label{sec:futurework}
We believe that apart from suggesting \jd{an interactively auditable ML pipeline,}\finalcut{such as  visual privacy aware ML pipeline,} future research should \finalcut{look for}\jd{consider} the trade-off between privacy, \finalcut{bias}\jd{fairness}, and \jd{model }accuracy in these systems.
In addition, we will develop a strategy for determining a failure rate for \jd{ML} models \jd{to provide a mechanism for researchers \cg{to} continually evaluate if their system is successful.}\finalcut{ and how those rates can be calculated, which could prove a vital area for future endeavors research.}
These \finalcut{suggestions}\jd{consideration}\cg{s} warrant further investigation to determine the success and limitations of \finalcut{prototyping}\jd{deployed systems}.

% What is ethical decision making for VP research?
% How can visual privacy research affect the surrounding communities?
% Experiment to study the trade off between privacy, bias and accuracy.
% further discuss trade offs and success
% Should researchers be looking to define a specific metric or point 
% at which they should halt deployment?
% What is a failure rate and how can it be calculated (if at all)?
%%%%%%%%%%%%%%%%%%%%%%%%%%%%%%%%%%%%%%%%%%%%%%%%%%%%%%%%%%%%%%%%%%%%%%%%%%%%%%%%%
\section{Acknowledgments}
The researchers are partially supported by awards from the Department of Defense SMART Scholarship and the National Science Foundation under Grant No. \#1952181.
% and the Oklahoma Louis Stokes Alliance for Minority Participation (OK-LSAMP).

\bibliographystyle{IEEEtran}
\bibliography{citations}

\end{document}